\documentclass{llncs}

\usepackage{makeidx}

\usepackage{amsmath}
\usepackage{amssymb}
\usepackage{stmaryrd}
\usepackage{pifont}
\usepackage[all]{xy}

\newcommand{\keywords}[1]{\par\addvspace\baselineskip
\noindent\keywordname\enspace\ignorespaces#1}

\begin{document}

\mainmatter              


\title{Yet Another Deep Embedding of \emph{B}:\\Extending \emph{de Bruijn} Notations}

\titlerunning{\emph{B} in \emph{Coq}}

\author{\'Eric Jaeger\inst{1}\inst{2} \and
        Th\'er\`ese Hardin\inst{1}}

\authorrunning{Jaeger, Hardin}


%
\institute{
LIP6, UPMC,
4 place Jussieu, 75252 Paris Cedex 05, France
\and
DCSSI,
51 boulevard de La Tour-Maubourg, 75700 Paris 07 SP, France
}

\maketitle


\newcommand{\tool}[1]{\emph{\small\sc#1}}
\newcommand{\Bicoqd}{\tool{BiCoq}\:}
\newcommand{\Bicoqt}{\tool{BiCoq$_3$}\:}

\newcommand{\NAT}{\ensuremath{\mathbb{N}}}

\newcommand{\choice}{\ensuremath{\talloblong}}
\newcommand{\guard}{\ensuremath{\Longrightarrow}}

\newcommand{\rulename}[1]{\ensuremath{\:{\scriptstyle{[{#1}]}}}}

\newcommand{\Bidx}{\ensuremath{\mathbb{I}}}
\newcommand{\dbzero}{\ensuremath{\text{0}}}
\newcommand{\dbone}{\ensuremath{\text{1}}}
\newcommand{\dbtwo}{\ensuremath{\text{2}}}
\newcommand{\dbthree}{\ensuremath{\text{3}}}
\newcommand{\dbfour}{\ensuremath{\text{4}}}
\newcommand{\dbfive}{\ensuremath{\text{5}}}

\newcommand{\Bnamp}{\ensuremath{\mathbb{K}}}
\newcommand{\Bnamo}{\ensuremath{\mathbb{J}}}

\newcommand{\Bbol}{\ensuremath{\mathbb{B}}}

\newcommand{\Bprd}{\ensuremath{\mathbb{P}}}
\newcommand{\Bexp}{\ensuremath{\mathbb{E}}}
\newcommand{\Btrm}{\ensuremath{\mathbb{T}}}
\newcommand{\pand}{\ensuremath{\dot{\land}}}
\newcommand{\pimp}{\ensuremath{\dot{\Rightarrow}}}
\newcommand{\pnot}{\ensuremath{\dot{\lnot}}}
\newcommand{\pfor}{\ensuremath{\underline{\forall}}}
\newcommand{\pequ}{\ensuremath{\dot{=}}}
\newcommand{\pins}{\ensuremath{\dot{\in}}}
\newcommand{\ppvr}[1]{\ensuremath{\dot{\pi}_{#1}}}
\newcommand{\evar}[1]{\ensuremath{\dot{\chi}_{#1}}}
\newcommand{\ecpl}{\ensuremath{\dot{\mapsto}}}
\newcommand{\echs}{\ensuremath{\dot{\Downarrow} \;\!}}
\newcommand{\epro}{\ensuremath{\dot{\times}}}
\newcommand{\epow}{\ensuremath{\dot{\Uparrow} \;\!}}

\newcommand{\ecmp}[2] {\ensuremath{\stackrel{\{}{\_}
                                   {#1}{\stackrel{|}{\_}}{#2}
                                   \stackrel{\}}{\_}}}

\newcommand{\ebig}{\ensuremath{\dot{\Omega}}}
\newcommand{\ebie}[1]{\ensuremath{\dot{\omega}_{#1}}}
\newcommand{\piff}{\ensuremath{\dot{\Leftrightarrow}}}
\newcommand{\por}{\ensuremath{\dot{\lor}}}
\newcommand{\pexs}{\ensuremath{\dot{\exists}}}

\newcommand{\Bnotfree}{\ensuremath{\dot{\smallsetminus}}}

\newcommand{\DBinter}[1]{\left\llbracket {#1} \right\rrbracket}

\newcommand{\fresh}{\ensuremath{\mathcal{F}}}
\newcommand{\depth}{\ensuremath{\mathcal{D}}}

\newcommand{\Cmd}[1]{\ensuremath{\mathbf{#1}}}

\newcommand{\Bbdfor}[2]{\ensuremath{\!\uparrow_{\!\forall\!}\!({#1}\!\cdot\!{#2})}}
\newcommand{\Bbdexs}[2]{\ensuremath{\!\uparrow_{\!\exists\!}\!({#1}\!\cdot\!{#2})}}
\newcommand{\Bbdcmp}[3]{\ensuremath{\!\uparrow_{\!\{\!\}}\!\!({#1}\!:\!{#2}\!\cdot\!{#3})}}
\newcommand{\Binfor}[2]{\ensuremath{\!\downarrow_{\forall\!}\!({#2}\!\gets\!{#1})}}
\newcommand{\Binexs}[2]{\ensuremath{\!\downarrow_{\exists\!}\!({#2}\!\gets\!{#1})}}
\newcommand{\Bincmp}[2]{\ensuremath{\!\downarrow_{\{\!\}}\!\!({#2}\!\gets\!{#1})}}
\newcommand{\Baffec}[3]{\ensuremath{[{#1}\!:=\!{#2}]{#3}}}

\newcommand{\Bmaf}{\ensuremath{\mathbb{M}}}
\newcommand{\Bmaffec}[2]{\ensuremath{\langle\!\langle{#1}|{#2}\rangle\!\rangle}}

\newcommand{\Baffprd}[3]{\ensuremath{\langle{#1}\!:\equiv\!{#2}\rangle{#3}}}
\newcommand{\Bgraftprd}[3]{\ensuremath{[{#1}\!\lhd\!{#2}]{#3}}}

\newcommand{\Binf}{\ensuremath{\:\dot{\vdash}\:}}

\spnewtheorem*{notation}{Notation}{\bfseries}{\itshape}
\spnewtheorem*{defn}{Definition}{\bfseries}{\itshape}


\begin{abstract}
We present \Bicoqt\!\!, a deep embedding of the \tool{B} system in \tool{Coq}, focusing on the
technical aspects of the development. The main subjects discussed are related to the
representation of sets and maps, the use of induction principles, and the introduction of a
new \emph{de Bruijn} notation providing solutions to various problems related to the
mechanisation of languages and logics.
\keywords{formal methods, deep embedding, \emph{de Bruijn} notation}
\end{abstract}


Embedding a language or a logic is now a well-established practice in the academic community,
to answer various types of concerns, \emph{e.g.} normalisation of terms and influence of
reduction strategies for a programming language or consistency for a logic. It indeed supports
such meta-theoretical analyses as well as comparing and promoting interesting concepts and
features of other languages, or developing mechanically checked tools to deal with a language.

But a lot of difficulties arise that have to be addressed. First of all, an important design
choice has to be made between \emph{shallow} and \emph{deep} approaches, consistently with the
objectives of the embedding. Justifying the validity of an embedding -- its correctness and
completeness -- can also be difficult. Finally, a lot of technical details have to be
considered \emph{e.g.} to manage variables.

We address these questions through the presentation of \Bicoqd and \Bicoqt\:\!\!\!, two
versions of a deep embedding of the \tool{B} logic in the \tool{Coq} system. The main
objective for these embeddings is to evaluate the correctness of the \tool{B} method itself,
in the context of security developments; other objectives include the development of proven
tools for the \tool{B} and the derivation of new results about the \tool{B} logic. Yet we
focus in this paper on the \emph{technical} aspects of these embeddings, and explain the need
for a full redevelopment between the two versions by describing painfully learned lessons.
The presentation includes the definition of an extended \emph{de Bruijn} notation with
interesting potentialities to solve some frequently encountered problems related to the
mechanisation of languages.

This paper is divided into 6 sections. Sections \ref{aboutcoq}-\ref{aboutb} briefly
introduce \tool{Coq}, the notion of embedding and \tool{B}. Section \ref{debruijn} presents
\emph{de Bruijn} notations. The technical aspect of the development of \Bicoqd and \Bicoqt are
described in Sec. \ref{bicoq}, considering in particular \emph{de Bruijn} context management,
induction principles, techniques to implement maps and new results obtained through an
extension of the \emph{de Bruijn} notation using namespaces. Section \ref{conclusion}
concludes and identifies further activities.

\section{About \emph{Coq}}\label{aboutcoq}

\tool{Coq} \cite{coq:1} is a proof assistant based on a type theory. It offers a higher-order
logical framework that allows for the construction and verification of proofs, as well as the
development and analysis of functional programs in a \tool{ML}-like language with 
pattern-matching. It is possible in \tool{Coq} to define values and types, including dependent 
types (\emph{i.e.} types that explicitly depend on values); types of sort $\Cmd{Set}$
represent  sets of computational values, while types of sort $\Cmd{Prop}$ represent logical
propositions. When defining an inductive type -- which is a least fixpoint -- associated 
structural induction principles are automatically generated.

For the intent of this paper, it is sufficient to see \tool{Coq} as allowing for the
manipulation of inductive sets of terms and inductive logical properties. Let's consider the
following standard example:
\[\small\begin{array}{l}
\Cmd{Inductive}\:\NAT\!:\!\Cmd{Set}\!:=\!0\!:\!\NAT\:|\:S\!:\!\NAT\!\to\!\NAT\\
\Cmd{Inductive}\:\text{even}\!:\!\NAT\!\to\!\Cmd{Prop}\!:=
ev_0\!:\!\text{even}\:0\:|\:
ev_{2}\!:\!\forall (n\!:\!\NAT),\:\text{even}\:n\!\to\!\text{even}\:S(S\:n)
\end{array}\]
The first line defines a type $\NAT$ which is the smallest set of terms stable by application
of the constructors $0$ and $S$. $\NAT$ is exactly made of the terms $0$ and
$S(\ldots S(0)\ldots)$ for any finite iteration; being well-founded, structural induction on
$\NAT$ is possible. The second line defines a family of \emph{logical types}: $even\:0$ is a
type inhabited by the term $ev_0$,  $even\:2$ is an other type inhabited by
$(ev_{2}\:0\:ev_0)$, and $even\:1$ is an empty type. The standard interpretation is that
$ev_0$ is a proof of the proposition $even\:0$ and that there is no proof of $even\:1$, that
is we have $\lnot(even\:1)$. The intuitive view of our example is that $\NAT$ is a set of
terms, and $even$ a predicate marking some of them, defining a subset of $\NAT$.

\section{Deep and Shallow Embeddings}\label{embedding}

\emph{Embedding} in a proof assistant consists in mechanizing a \emph{guest} logic by encoding 
its syntax and semantic into a \emph{host} logic (\cite{gor:2,bou:1,azu:1}). In a
\emph{shallow} embedding, the encoding is partially based on a direct translation of the guest
logic into constructs of the host logic; in terms of programming languages, a shallow
embedding can intuitively be seen as the development of a translation function between two
langages, that is a compiler. On the contrary, a \emph{deep} embedding is better intuitively
described as the development of a virtual machine: the syntax and the semantic of the guest
logic are formalised as datatypes of the host logic. Taking the view presented in Sec.
\ref{aboutcoq}, the deep embedding of a logic defines the set of all sequents -- the terms --
and the subset of provable sequents (the inference rules of the guest logic being encoded as
constructors of the provability predicate).

Both approaches have pros and cons. The one we are concerned with, and that has led us to
choose the deep embedding approach, is \emph{accuracy}: a deep embedding allows for an exact
representation of the syntax and semantic of the guest logic, whereas a shallow embedding
appears to enforce a form of interpretation whose validity can be difficult to justify.

\section{About \emph{B}}\label{aboutb}

\subsection{A Short Description of \emph{B}}\label{aboutb_description}

\tool{B} \cite{abr:1} is a popular formal method that allows for the derivation of correct 
programs from specifications. Several industrial implementations are available (\emph{e.g.} 
\tool{AtelierB}, \tool{B Toolkit}), and it is widely used by both the academic world and the
industry for projects where  safety or security is mandatory.

The \tool{B} method defines a first-order predicate logic completed with elements of set
theory, a \emph{Generalised Substitution Language} (\tool{GSL}) and a methodology of
development based on the explicit concept of \emph{refinement}.

The logic is used to express preconditions, invariants, etc. and to conduct proofs. This logic
is not typed; a kind of well-formedness checking is described but is not integrated within the
logic.

The \tool{GSL} allows for the definitions of a form of \emph{Hoare} substitutions
\cite{hoa:1,dij:1,hoa:3} that can be abstract, declarative and non-deterministic (\emph{i.e.}
specifications) as well as concrete, imperative and deterministic  (\emph{i.e.} programs): the
substitution
$\Cmd{ANY}\:x\:\Cmd{WHERE}\:x^2\leq\!n\!<\!(x\!+\!1)^2$ for example specifies
$x\!\gets\!\sqrt{n}$.

Regarding the methodology, \tool{B} developments are made of \emph{machines} (modules
combining a \emph{state} in the form of variables, \emph{invariants} and \emph{operations}
described as generalised substitutions to read or alter the state). Intuitively a machine
$M_C$ \emph{refines} a machine $M_A$ if any observable behaviour of $M_C$ is a possible
behaviour of $M_A$ -- this encompasses the notion of correctness. Refinement being transitive,
it is possible to go progressively from the specification to the implementation; by
discharging at each step the \emph{proof obligations} of the \tool{B} method, a program can be
proven to be a correct and complete implementation of a specification.

Note that the language represented by the \tool{GSL} is imperative; at the last stage of
refinement the machines are written using only the \tool{B0} sublanguage of the \tool{GSL} and
are easily translated \emph{e.g.} into \tool{C} programs.

\subsection{Embedding \emph{B}: Related Works and Motivations}\label{aboutb_motivations}

Shallow embeddings of \tool{B} in higher-order logics have been proposed in several papers
(cf. \cite{bod:1b,cha:1}) formalising the \tool{GSL} in \tool{PVS}, \tool{Coq} or
\tool{Isabelle/HOL}. Such embeddings are not dealing with the \tool{B} logic, and by using 
directly the host logic to express \tool{B} notions, they introduce a form of interpretation
-- which is fully acceptable for example to promote the \tool{B} methodology in other formal
methods.

The objectives of \Bicoqd and \Bicoqt are very different, the main concern being related to
validation. Indeed, the \tool{B} method is used for the development of safe or secure systems
(\emph{e.g.} \cite{beh:1,bie:1}), and it is therefore important to know what is the level of
confidence that one can grant to a system proven using this method, and how to improve this
level of confidence. The other objectives are the development of formally checked tools for
\tool{B} developments, illustrated by a proven prover (not discussed further in this paper but
detailed in \cite{jae:1}) and the derivation of new results about the \tool{B} logic.
Regarding the latter, it is again important to be able to justify that such results are not a
consequence of the embedding itself, \emph{e.g.} using an `alien' trick provided by
\tool{Coq}, and are indeed valid for use in a standard \tool{B} development.

With the objectives of accuracy and independancy, the translation for a shallow embedding
would be difficult to define but also to defend against a skeptical independent evaluator.
Consider \tool{B} functions that are relations, possibly partial and undecidable: translating
accurately this concept in \tool{Coq} is a tricky exercise. A deep embedding makes the
justification easier, and has also the advantage to clearly separate the host and the guest
logics: excluded middle, provable in the \tool{B} logic as well as in \Bicoqd or
\Bicoqt\:\!\!\!, is not promoted to the \tool{Coq} logic. Such a deep embedding of the
\tool{B} logic in \tool{Coq} is described in \cite{brk:1}, to validate the \emph{base rules}
used by the prover of \tool{AtelierB} -- yet not checking standard \tool{B} results, and
without implementation goal.

\section{\emph{De Bruijn} Notations}\label{debruijn}

There are numerous problems to deal with when mechanising a language (cf. \cite{pop:1,ayd:1}),
one of them being related to the representation of bound variables. Indeed, two terms
differing only by the names of their bound variables ($\alpha$-renaming), such as
$\lambda x\cdot\lambda y\cdot x-y$ and $\lambda z\cdot\lambda x\cdot z-x$, should be
considered as equal but are not when using a notation with names (denoted $\lambda_{\text{V}}$
in this paper); one may also wonder how to compute the reduction of the substitution
$[x\!:=\!E]\lambda x\!\cdot\!T$.

\emph{De Bruijn} notations (cf. \cite{deb:1,gor:1} or more recently \cite{nor:1}) address
these problems by encoding bound variables as natural values pointing to a binder; they define
an $\alpha$-quotiented representation, \emph{i.e.} terms equivalent modulo $\alpha$-renaming
are indeed equal. They also provide a clear semantic to deal with capture phenomena applicable
between others when considering substitutions.

\subsection{\emph{De Bruijn} Indexes: The $\lambda_{\text{dBi}}$ Notation}
\label{debruijn_indexes}

The most known \emph{de Bruijn} notation uses indexes, that are relative pointers counting
binders from the variables (the leaves in the tree representing the term). The value $0$
represents the variable bound by the closest parent binder, as illustrated hereafter
(\emph{de Bruijn} binders are underlined for the sake of clarity):
\[\small\begin{array}{lll}
\lambda_{\text{V}}\text{ notation}& \quad \quad &
\lambda x\!\cdot\!\lambda y\!\cdot\!(X_0\!+\!x\!-\!y)\\
\lambda_{\text{dBi}}\text{ notation} &&
\underline{\lambda}\underline{\lambda}(2\!+\!1\!-\!0)
\end{array}\]
We have chosen here to use the \emph{pure nameless notation}: the free variable $X_0$ is
represented by the value $2$, assuming it is the first free variable in the context (left
implicit here). Such a pointer is said to be \emph{dangling} as its value exceeds the number
of parent binders. Another possible alternative is to use the
\emph{locally nameless notation}; in this case, free variables are represented by names
(and are syntactically different of bound variables). We will not consider further this
approach that requires to give a specific semantic to dangling pointers or to manage side
conditions enforcing terms to be ground (without dangling pointers).

\subsection{\emph{De Bruijn} Levels: The $\lambda_{\text{dBl}}$ Notation}
\label{debruijn_levels}

Another option when defining a \emph{de Bruijn} representation is to use levels, discussed
\emph{e.g.} in \cite{hol:1}. Levels are absolute pointers counting binders from the root of
the term; the value $0$ then represents the variable bound by the farest parent binder, as
illustrated here:
\[\small\begin{array}{lll}
\lambda_{\text{V}}\text{ notation} & \quad \quad &
\lambda x\!\cdot\!\lambda y\!\cdot\!(X_0\!+\!x\!-\!y)\\
\lambda_{\text{dBl}}\text{ notation} &&
\underline{\lambda}\underline{\lambda}(2\!+\!0\!-\!1)
\end{array}\]
Index and level notations only differ in the representation of bound variables. Levels ensure
a unique representation in a term of a bound variable, whereas with indexes this
representation depends on the variable position; on the other hand, bound levels need frequent
renumbering during abstraction or substitution whereas bound indexes are never modified. Other
pros and cons of these approaches will be considered later in the paper to explain \Bicoqt
design choices.

\subsection{Managing \emph{de Bruijn} Indexes in $\lambda$-Calculus}
\label{debruijn_managing}

As mentioned, the index representing a given bound variable change with its
\emph{$\lambda$-height}, \emph{i.e.} the number of parent binders, as illustrated by this
example:
\[\small\begin{array}{lll}
\lambda_{\text{V}}\text{ notation} & \quad \quad &
\lambda x\!\cdot\!(x\!+\!\lambda y\!\cdot\!(x\!-\!y) X_0)\\
\lambda_{\text{dBi}}\text{ notation} &&
\underline{\lambda}(0\!+\!\underline{\lambda}(1\!-\!0) 2)
\end{array}\]
This makes manipulating $\lambda_{\text{dBi}}$ terms by hand rather awkward. It is therefore
customary to provide standard operators to support index management, either technical such as
\emph{lifting} or user-relevant such as \emph{substitution}. The former is used by the latter
to adapt terms when crossing a binder, as illustrated here (where $\Btrm$ denotes the set of
$\lambda_{\text{dBi}}$ terms, $i$ an index in $\Bidx\!=\!\NAT$, $\uparrow$ the lifting and
$\Baffec{i}{E}{T}$ the replacement of all occurences of the \emph{free} variable $i$ in $T$ by
$E$):
\[\small\begin{array}{lll}
 \begin{array}{l}
  \uparrow_d:\!\Btrm\!\to\!\Btrm\!:=\\
  \quad|\:\underline{\lambda} T'\Rightarrow
          \underline{\lambda}(\uparrow_{d\!+\!1} T')\\
  \quad|\:i'\Rightarrow
          \Cmd{if}\:d\!\leq\!i'\:\Cmd{then}\:i'\!+\!1\:\Cmd{else}\:i'\\
  \quad|\:\ldots
 \end{array}
& \quad \quad \quad &
 \begin{array}{l}
  \Baffec{i}{E}{}\!:\!\Btrm\!\to\!\Btrm\!:=\\
  \quad|\:\underline{\lambda} T'\Rightarrow
          \underline{\lambda}(\Baffec{x\!+\!1}{\uparrow\!E}{T'})\\
  \quad|\:i'\Rightarrow
          \Cmd{if}\:i\!=\!i'\:\Cmd{then}\:E\:\Cmd{else}\:i'\\
  \quad|\:\ldots
 \end{array}
\end{array}\]
Indeed, crossing a binder modifies the $\lambda$-height, so the index $i$ has to be
incremented to represent the same variable, and similarly \emph{dangling} indexes of $E$ have
to be incremented to maintain their semantic as well as to avoid their capture -- this is the
role of lifting. To identify dangling indexes, lifting is parameterised by the
\emph{contextual information} $d$ recording the current $\lambda$-height, left implicit when
$d\!=\!0$ (other values of $d$ resulting only from recursive calls for bound subterms).

This toolbox for $\lambda$-calculus is completed with operators defining a user-friendly
representation, as in \cite{gor:1}. The idea is to emulate the $\lambda_\text{V}$ abstraction,
a not so simple transformation in $\lambda_\text{dBi}$ as illustrated here (capturing $X_1$):
\[\small\begin{array}{lllll}
\lambda_{\text{V}}\text{ notation} & \quad \quad &
X_0\!+\!X_1\!+\!X_2 &
\quad\to\quad &
\lambda x\!\cdot\!(X_0\!+\!x\!+\!X_2)
\\
\lambda_{\text{dBi}}\text{ notation} &&
0\!+\!1\!+\!2 &
\quad\to\quad &
\underline{\lambda}(1\!+\!0\!+\!3)
\end{array}\]
To this end, we define the abstraction function
$\lambda(i\!\cdot\!T)\!:=\!\underline{\lambda}(\text{Abstr}_0\:i\:T)$ with:
\[\small\begin{array}{l}
 \text{Abstr}_d(i\!:\!\Bidx)\!:\!\Btrm\!\to\!\Btrm\!:=\\
 \quad|\:\underline{\lambda} T'\Rightarrow
       \underline{\lambda}(\text{Abstr}_{d+1}\:(i\!+\!1)\:T')\\
 \quad|\:i'\!\Rightarrow\!\left\{
         \begin{array}{l}
         i'\text{ if }i'<d\\
         d\text{ if }i'\geq d \text{ and } i'=i\\
         i'\!+\!1\text{ if }i'\geq d \text{ and } i'\not=i\\
         \end{array}\right.\\
 \quad|\:\ldots
\end{array}\]
Here $\lambda(i\!\cdot\!T)$ is not the $\lambda_\text{V}$ abstraction but a \emph{function}
computing the correct $\lambda_\text{dBi}$ term, defining a \emph{form} of $\lambda_\text{V}$
representation ($i$ being an index and $T$ a $\lambda_\text{dBi}$ term).

\section{A Detailed presentation of \emph{BiCoq3}}\label{bicoq}

We now discuss the design choices made for developing \Bicoqt\:\!\!\!, also addressing the
technical alternatives and their consequences. From this point, illustrations and codes will
describe the \tool{B} logic as encoded in \tool{Coq}, instead of the $\lambda$-calculus
considered up to now; dotted notations will represent \tool{B} logical operators in \tool{Coq}
(\emph{e.g.} $\lnot$ is the \tool{Coq} negation and $\dot{\lnot}$ the embedded \tool{B}
negation).

\subsection{Embedding the Syntax}\label{bicoq_syntax}

\subsubsection{Using \emph{de Bruijn} indexes.} We have chosen for \Bicoqd and \Bicoqt to use
a \emph{de Bruijn} notation, and have investigated both indexes and levels: two \emph{full}
versions of \Bicoqt have been developed, yet without reaching a general conclusion. Indeed for
\emph{most} of our needs, levels are more efficient; they are easier to deal with, theorems
tend to be more generic and proofs simpler. Consider as a typical example the lifting
functions for indexes (left code) and levels (right code):
\[\small\begin{array}{lll}
 \begin{array}{l}
  \uparrow_d:\!\Btrm\!\to\!\Btrm\!:=\\
  \quad|\:\underline{\lambda} T'\Rightarrow
          \underline{\lambda}(\uparrow_{d\!+\!1} T')\\
  \quad|\:i'\Rightarrow
          \Cmd{if}\:d\!\leq\!i'\:\Cmd{then}\:i'\!+\!1\:\Cmd{else}\:i'\\
  \quad|\:\ldots
 \end{array}
& \quad \quad \quad &
 \begin{array}{l}
  \uparrow^\text{L}:\!\Btrm\!\to\!\Btrm\!:=\\
  \quad|\:\underline{\lambda} T'\Rightarrow\underline{\lambda}(\uparrow^\text{L} T')\\
  \quad|\:i'\!\Rightarrow\!i'\!+\!1\\
  \quad|\:\ldots
 \end{array}
\end{array}\]
As mentioned in Sub. \ref{debruijn_managing}, $\uparrow_d$ requires a contextual parameter to
identify dangling indexes, bound indexes being never modified. On the contrary its
$\lambda_{\text{dBl}}$ equivalent $\uparrow^\text{L}$ increments all levels, so this parameter
is not required and theorems about lifting are not specialised according to its value.

Our final (and late) choice is however to use \emph{de Bruijn} indexes. Indeed complex results
in our developement require as a proof tool the definition of
\emph{parallel $\lambda$-substitutions} providing an alternative encoding of standard
operations on terms (such as lifting). This is feasible with $\lambda_{\text{dBi}}$, those
operations being similar to substitutions in never modifying bound indexes, but not in
$\lambda_{\text{dBl}}$. We therefore consider that whereas \emph{de Bruijn} levels are simpler
to use, there is a clear advantage for \emph{de Bruijn} indexes when dealing with
\emph{advanced} techniques related \emph{e.g.} to term transformations under binders detailed
later in this paper.

\subsubsection{Representing \emph{B} terms.} Given a set of identifiers $I$, the \tool{B}
logic syntax defines predicates $P$, expressions $E$, sets $S$ and variables $V$ as follows:
\[\small\begin{array}{rclclclclclclclcl}
\rule{1.0cm}{0cm} &
\rule{0.7cm}{0cm} & \rule{0.8cm}{0cm} &
\rule{0.2cm}{0cm} & \rule{0.8cm}{0cm} &
\rule{0.2cm}{0cm} & \rule{0.8cm}{0cm} &
\rule{0.2cm}{0cm} & \rule{0.8cm}{0cm} &
\rule{0.2cm}{0cm} & \rule{0.8cm}{0cm} &
\rule{0.2cm}{0cm} & \rule{0.8cm}{0cm} &
\rule{0.2cm}{0cm} & \rule{0.8cm}{0cm} &
\rule{0.2cm}{0cm} & \rule{0.8cm}{0cm} \vspace{-12pt} \\
P &:=& P\!\land\!P &|& P\!\!\Rightarrow\!\!P &|& \lnot P &|& \forall\:V\cdot P
        &|& E\!=\!E &|& E\!\in\!S &|& [V\!\!:=\!\!E]P 
\\
E &:=& V &|& S &|& E\!\mapsto\!E  &|& \Downarrow S &|& [V\!\!:=\!\!E]E 
\\
S &:=& \Cmd{BIG} &|& \Uparrow S &|& S\!\times\!S  &|& \{V|P\} 
\\
V &:=& I &|& V,V
\end{array}\]
In this syntax, $[V\!:=\!E]T$ represents the (elementary) substitution, $V_1,V_2$ a list 
of variables, $E_1\!\mapsto\!E_2$ a pair of expressions, $\!\Downarrow\!$ and  $\!\Uparrow\!$ 
the \emph{choice} and \emph{powerset} operators, and $\Cmd{BIG}$ a constant set. Other
connectors are standard, and new connectors are defined from the previous ones,
$P\!\!\Leftrightarrow\!\!Q$ as $P\!\!\Rightarrow\!\!Q \land Q\!\!\Rightarrow\!\!P$,
$P\!\lor\!Q$ as $\lnot P\!\!\Rightarrow\!\!Q$, $\exists\:V\!\cdot\!P$ as 
$\lnot\forall\:V\!\cdot\!\lnot P$, $S\!\subseteq\!T$ as $S\!\in\:\Uparrow\!T$, etc.

The \tool{B} syntax is formalised in \tool{Coq} by two mutually inductive types with the
following constructors\footnote{This is a slightly simplified presentation of \Bicoqt focusing
on relevant aspects.}, $\Bidx$ being the set of indexes (\emph{i.e.} $\NAT$):
\[\small\begin{array}{rclclclclclclclcl}
\rule{1.0cm}{0cm} &
\rule{0.7cm}{0cm} & \rule{0.8cm}{0cm} &
\rule{0.2cm}{0cm} & \rule{0.8cm}{0cm} &
\rule{0.2cm}{0cm} & \rule{0.8cm}{0cm} &
\rule{0.2cm}{0cm} & \rule{0.8cm}{0cm} &
\rule{0.2cm}{0cm} & \rule{0.8cm}{0cm} &
\rule{0.2cm}{0cm} & \rule{0.8cm}{0cm} &
\rule{0.2cm}{0cm} & \rule{0.8cm}{0cm} &
\rule{0.2cm}{0cm} & \rule{0.8cm}{0cm} \vspace{-12pt} \\
\Bprd &:=& \Bprd \pand \Bprd &|& \Bprd \pimp \Bprd &|& \pnot \Bprd &|& \pfor \Bprd
        &|& \Bexp \pequ \Bexp &|& \Bexp \pins \Bexp
\\
\Bexp &:=& \evar{}\Bidx &|& \Bexp \ecpl \Bexp &|& \echs \Bexp  &|& \ebig &|& \epow \Bexp
       &|& \Bexp \epro \Bexp &|& \ecmp{\Bexp}{\Bprd}
\end{array}\]
$\Bprd$ represents \tool{B} predicates and $\Bexp$ merges \tool{B} expressions $E$, sets $S$
and variables $V$ to enrich the \tool{B} syntax that is too strict (\emph{e.g.}
$E\!\!\in\Downarrow\!\!(\Uparrow\!S)$ is syntactically invalid in standard \tool{B}). In the
rest of this paper $\Btrm\!=\!\Bprd\!\cup\!\Bexp$ denotes the type of \tool{B} terms.

$\ebig$ represents the constant set $\Cmd{BIG}$, $\evar{}$ unary \emph{de Bruijn} variables
(using $\evar{i}$ to denote the application of constructor $\evar{}$ to $i\!:\!\Bidx$). The
\emph{B} binders $\forall V\!\cdot\!P$ and $\{V|P\}$ are respectively represented by the
constructors $\pfor$ and $\ecmp{\!}{\!}$, that are raw \emph{de Bruijn} binders (we therefore
use the underlined notation, the dotted notation $\dot{\forall}$ and $\dot{\{}\dot{|}\dot{\}}$
being reserved for a user-friendly notation, cf. Sub. \ref{debruijn_managing}). Using
\emph{de Bruijn} indexes, they have no attached names and only bind a single variable --
binding over list of variables being eliminated without loss of expressivity\footnote{Remark
by the way that the notation
$\{V_1\;\!\!,\;\!\!V_2\:|\:V_1\;\!\!,\;\!\!V_2\!\in\!S_1\!\times\!S_2\!\land\!P\}$ used in
\cite{abr:1} is an example of syntactically invalid term confusing the expression 
$x\!\mapsto\!y$ with the variable $x\;\!\!,\;\!\!y$, whose `correct' version
$\{V_1\;\!\!,\;\!\!V_2\:|\:V_1\!\mapsto\!V_2\!\in\!S_1\!\times\!S_2\!\land\!P\}$ is not
well-formed.}. The constructor $\ecmp{\!}{\!}$ is further modified to keep in the syntax
definition only well-formed terms (cf. Sub. \ref{aboutb_description}). Indeed, the
well-formedness checking in \tool{B} requires comprehension sets to be of the form
$\{x\:|\:x\!\in\!S\land P\}$ with $x$ not free in $S$. \emph{Both} constraints are embedded in
our syntax. The comprehension set constructor has two parameters, the left one being an
expression representing $S$ and the right one a predicate representing $P$; the non-freeness
condition is ensured by considering this constructor as a binder only for its right
parameter\footnote{Similarly consider the $\lambda x\!:\!T\!\cdot\!E$ notation in simply-typed
$\lambda$-calculus; the $\lambda$ captures $x$ in $E$ but not in $T$, binding only one of its
parameters.}. This bridges the gap between syntactically correct terms and well-formed ones.

Note finally that we do not represent \tool{B} syntactical constructs $[V\!\!:=\!\!E]T$
(elementary substitutions); this will be justified later in this paper.

\subsection{\emph{De Bruijn} Management: Improving Context Awareness}\label{bicoq_context}

We ease the use of the \emph{de Bruijn} notations by providing functions, as in Sub.
\ref{debruijn_managing}. First of all, lifting is adapted to our constructors -- noting that
as $\ecmp{}{}$ does not bind its left parameter, the left $\lambda$-height is not incremented:
\[\small\begin{array}{rclcl}
\uparrow_d:\!\Btrm\!\to\!\Btrm
& :=  & \pfor P' & \Rightarrow & \pfor (\uparrow_{d\!+\!1} P')\\
& \:| & \ecmp{E'}{P'} & \Rightarrow & \ecmp{\uparrow_d E'}{\uparrow_{d\!+\!1} P'}\\
& \:| & \evar{i'} & \Rightarrow &
        \evar{}(\Cmd{if}\:d\!\leq\!i'\:\Cmd{then}\:i'\!+\!1\:\Cmd{else}\:i')\\
& \:| & \ldots & \Rightarrow & \ldots\text{(straightforward recursion)}
\end{array}\]
We also define abstraction functions, but with \emph{additional} subtle changes:
\[\small\begin{array}{rclcl}
\text{Abstr}_d(i\!:\!\Bidx)\!:\!\Btrm\!\to\!\Btrm
& :=  & \pfor P' & \Rightarrow & \pfor (\text{Abstr}_{d\!+\!1}\:(\uparrow_d i)\:P')\\
& \:| & \ecmp{E'}{P'} & \Rightarrow & \ecmp{\text{Abstr}_{d}\:i\:E'\:}
                                           {\:\text{Abstr}_{d\!+\!1}\:(\uparrow_d i)\:P'}\\
& \:| & \evar{i'} & \Rightarrow &
        \evar{}(\Cmd{if}\:i\!=\!i'\:\Cmd{then}\:d\:\Cmd{else}\uparrow_d i)\\
& \:| & \ldots & \Rightarrow & \ldots\text{(straightforward recursion)}
\vspace{3pt}
\\
\multicolumn{5}{l}
{
\begin{array}{lllll}
\dot{\forall}\:i\!\cdot\!P\!:=\!\pfor(\text{Abstr}_0\:i\:P)
& \quad &
\dot{\exists}\:i\!\cdot\!P\!:=\!\pnot(\dot{\forall}i\!\cdot\!\pnot P)
& \quad &
\dot{\{}i\!:\!E\:\dot{|}\:P\dot{\}}\!:=\ecmp{E\:}{\:\text{Abstr}_0\:i\:P}
\end{array}
}
\end{array}\]
Compared with the abstraction function defined in Sub. \ref{debruijn_managing}, it is
important to note the difference w.r.t. the $\lambda$-height parameter $d$. We do not
\emph{increment} indexes anymore but we \emph{lift} them; furthermore when we lift an
expression, we ensure that we use $d$ instead of the default value $0$. This does not change
the result: applying $n$ times the function $\!\uparrow_0$ yields exactly the same result as
applying successively $\uparrow_0$, $\uparrow_1$, \ldots, $\uparrow_{n\!-\!1}$ -- benefits are
not \emph{computational} but \emph{logical}. Indeed we have (painfully) discovered that a
stricter discipline in managing contexts is a very good practice, easing the expression of
theorems as well as their proofs. In fact, this discipline leads to generalise the
$\lambda$-height parameter to functions \emph{that don't need it}. For example, deciding if a
variable appears free in a term does not require this parameter (left code), but proofs are
easier by adding it \emph{and} using it to lift the variable parameter (right code):
\[\small\begin{array}{ll}
 \begin{array}{l}
  \text{Free}(i\!:\!\Bidx)\!:\!\Btrm\!\to\!\Bbol\!:=\\
  \quad|\:\pfor P'\Rightarrow \text{Free}\:(i\!+\!1)\:P'\\
  \quad|\ecmp{E'}{P'}\Rightarrow \text{Free}\:i\:E' \lor \text{Free}\:(i\!+\!1)\:P'\\
  \quad|\:\evar{i'}\Rightarrow i'\!=\!i\\
  \quad|\:\ldots\Rightarrow\ldots\text{(straightforward recursion)}
 \end{array}
&
 \begin{array}{l}
  \text{Free}_d(i\!:\!\Bidx)\!:\!\Btrm\!\to\!\Bbol\!:=\\
  \quad|\:\pfor P'\Rightarrow \text{Free}_{d\!+\!1}\:(\uparrow_d i)\:P'\\
  \quad|\ecmp{E'}{P'}\Rightarrow
        \text{Free}_d\:i\:E' \lor \text{Free}_{d\!+\!1}\:(\uparrow_d i)\:P'\\
  \quad|\:\evar{i'}\Rightarrow i'\!=\!i\\
  \quad|\:\ldots\Rightarrow\ldots\text{(straightforward recursion)}
 \end{array}
\end{array}\]
Generalising the $\lambda$-height parameter and using it ensures an explicit management of the
context, a form of weak typing useful for complex proofs.

We also define additional functions (not described in Sub. \ref{debruijn_managing}) to deal
with the \tool{B} syntactical constructs $[V\!\!:=\!\!E]T$ not represented in our syntax.
It is our view that these constructs are introduced early in \tool{B} \emph{only} for
expressing inference rules such as the $\forall$-elimination
($\Gamma\!\vdash\!\forall V\!\cdot\!P\!\to\!\Gamma\!\vdash\![V\!:=\!E]P$), that is a form of
application followed by $\beta$-reduction as in standard $\lambda$-calculus; there is no
reason to enforce this operation to be the \tool{B} elementary substitution defined by the
\tool{GSL}... Neither do we represent the application in our syntax, as in standard
formalisations of $\lambda$-calculus: representing application (and $\beta$-reduction either
as an external or internal operation \emph{e.g.} using the \emph{explicit substitution}
approach \cite{aba:1,cur:1}) is interesting for example to study normalisation strategies, but
this is not relevant in our case. We encode \emph{directly} such elimination rules,
\emph{i.e.} application followed by $\beta$-reduction, as an external operation, through
\emph{application functions} in \tool{Coq} denoted $T@_{\forall}E$ and $T@_{\{\!\}}E$, one per
binder\footnote{These functions only apply to terms starting with the appropriate binder;
the partiality is encoded in \tool{Coq} by an additonal proof parameter left implicit here.}:
\[\small\begin{array}{rclcl}
\text{App}_d(E\!:\!\Bexp)\!:\!\Btrm\!\to\!\Btrm
& :=  & \pfor P' & \Rightarrow & \pfor (\text{App}_{d\!+\!1}\:(\uparrow_d E)\:P')\\
& \:| & \ecmp{E'}{P'} & \Rightarrow & \ecmp{\text{App}_{d}\:E\:E'\:}
                                           {\:\text{App}_{d\!+\!1}\:(\uparrow_d E)\:P'}\\
& \:| & \evar{i'} & \Rightarrow &
        \left\{
          \begin{array}{l}
          \evar{i'-1}\text{ if }d<i'\\
          E\text{ if }d\!=\!i'\\
          \evar{i'}\text{ if }d>i'\\
          \end{array}\right.
\\
& \:| & \ldots & \Rightarrow & \ldots\text{(straightforward recursion)}
\vspace{3pt}
\\
\multicolumn{5}{l}
{T@_{\forall}E\!:=\!\Cmd{match}\:T\:\Cmd{with}\:\pfor\:T'\Rightarrow\text{App}_0\:E\:T'}
\\
\multicolumn{5}{l}
{T@_{\{\!\}}E\!:=\!
 \Cmd{match}\:T\:\Cmd{with}\:\ecmp{\!E'}{T'\!}\Rightarrow E\pins E'\pand\text{App}_0\:E\:T'}
\end{array}\]
The $\forall$-elimination can then be written
$\Gamma\!\vdash\!\forall V\!\cdot\!P\!\to\!\Gamma\!\vdash\!(\forall V\!\cdot\!P)@_{\forall}E$.
As abstraction, application and substitution functions are such that the following properties
hold (the left one being valid only after generalising the $\lambda$-height parameter to the
substitution function), our rule is equivalent to the standard one:
\[\small\begin{array}{l}
\Baffec{i}{E}{}_d T\!=\!\text{App}_d\:E\:(\text{Abstr}_d\:i\:T)
\quad \text{or more simply} \quad
\Baffec{i}{E}{T}=(\dot{\forall} i\!\cdot\!P)@_{\forall}E
\end{array}\]
The point is that we do not consider substitution as primitive. The standard definition of
$\beta$-reduction $\lambda x\!\cdot\!T@E\!\to_\beta\![x\!:=\!E]T$ describes the semantic of
application using substitution; in \Bicoqt on the contrary application is directly defined and
the substitution is a composite operation. Note also that we can write
$\text{App}_d\:\evar{i}\:(\text{Abstr}_d\:i\:T)\!=\!T$, or more simply
$(\dot{\forall} i\!\cdot\!P)@_{\forall}\evar{i}\!=\!T$, to emphasise that application is the
reverse of abstraction\footnote{This result commutes,
$\text{Abstr}_d\:i\:(\text{App}_d\:\evar{i}\:T)\!=\!T$ provided that
$\text{Free}_{d}\:i\:\underline{\forall}T\!=\!\bot$.}.

\subsection{Embedding the Inference Rules}

Having formalised the \tool{B} syntax as a datatype, the next step is to encode the \tool{B}
inference rules as the constructors of an inductive \emph{provability} predicate defining a
dependent type. We denote $\Gamma\Binf P$ the \tool{Coq} type of all \tool{B} proofs of $P$
under the assumptions $\Gamma$; if it is inhabited then $P$ is provable assuming $\Gamma$.
Note that $\lnot(\Gamma\Binf P)$, \emph{i.e.} `$\Gamma\Binf P$ is an empty type', is different
from $\Gamma\Binf\pnot P$.

Thanks to the use of the user-friendly functions described in Subs. \ref{debruijn_managing}
and \ref{bicoq_context}, the constructors look very much like the standard \tool{B}
rules\footnote{We also benefit from the \Cmd{Notation} command provided by \tool{Coq} to use
\tool{UTF-8} symbols instead of constructors or functions names.}. The translation is
straightforward, merely a syntactical one, limiting the risk of error, as illustrated here
(where $V\backslash\Gamma$ means that $V$ does not appear free in $\Gamma$):
\[\small\begin{array}{rcl}
\begin{array}{c}
\Gamma \vdash P \quad \Gamma \vdash Q \\\hline \Gamma \vdash P \land Q
\end{array}
& \quad\text{is encoded by}\quad &
\Gamma\Binf P\!\to\!\Gamma\Binf Q\!\to\!\Gamma\Binf P \pand Q
\\
\begin{array}{c}
\Gamma \vdash P \quad V \backslash \Gamma \\\hline
\Gamma \vdash \forall\: V \cdot P
\end{array}
& &
i \Bnotfree \Gamma\!\to\!\Gamma \Binf P\!\to\!\Gamma
\Binf \dot{\forall}\:i\!\cdot\!P
\end{array}\]
The main divergence is a correction of the definition of the cartesian product. Indeed, beyond
minor syntactical problems, \Bicoqd has also pointed out \tool{B} logical oversights; analyses
have shown that the following results, presented in \cite{abr:1} as theorems, are in fact not
provable with the standard \tool{B} inference rules\footnote{Further details are discussed in
\cite{jae:1}.}:
\[\small\begin{array}{l}
\vdash E_1\!\!\mapsto\!\!F_1\!=\!E_2\!\!\mapsto\!\!F_2\Rightarrow
 E_1\!=\!E_2 \land F_1\!=\!F_2\\
\vdash S_1\!\subseteq\!S_2 \land T_1\!\subseteq\!T_2 \Rightarrow
 S_1\!\times\!T_1\!\subseteq\!S_2\!\times\!T_2
\end{array}\]
To our knowledge, this was not known by the \tool{B} community -- whereas implementations of
the \tool{B} method correct this flaw, consciously or not. The flawed rule
$\vdash\!(E\!\!\mapsto\!\!F)\!\in\!(S\!\!\times\!\!T) \Leftrightarrow 
(E\!\!\in\!\!S)\!\land\!(F\!\!\in\!\!T)$ has therefore been replaced in \Bicoqd by:
\[\small\begin{array}{l}
\Gamma\Binf E_1\ecpl E_2\pequ E_3\ecpl E_4\!\to\!
 \Gamma\Binf E_1\pequ E_3 \pand \Binf E_2\pequ E_4\\
i_1,i_2 \Bnotfree E\pins(E_1\epro E_2)\!\to\!i_1 \neq i_2\!\to\!
  \Gamma\Binf\!\dot{\exists}\:i_1\!\cdot\! i_1\pins E_1\pand
    \dot{\exists}\:i_2\!\cdot\!i_2\pins E_2\pand E\pequ i_1\ecpl i_2\piff E\pins(E_1\epro E_2)
\end{array}\]

\subsection{A Generic Induction Principle}\label{bicoq_induction}

The definition of an inductive datatype in \tool{Coq} yields automatically the associated
structural induction principle. This principle is relevant to prove structural properties such
as those about freeness, but not to prove \emph{semantical} results.

Indeed, it identifies $T$ as the predecessor of $\pfor T$, \emph{i.e.} that proving $P(T)$ by
structural induction requires proving a subgoal of the form $P(T')\!\Rightarrow\!P(\pfor T')$.
But using \emph{de Bruijn} indexes this approach is not appropriate:
\[\small\begin{array}{llccc}
\text{\emph{de Bruijn} indexes} & \quad &
\underline{\exists}(1\!*\!0\!>\!2) & \quad &
\underline{\forall}(\underline{\exists}(1\!*\!0\!>\!2))
\\
\text{Natural notation} & \quad &
\exists\:z\!\cdot\!X_0\!*\!z\!>\!X_1 & \quad &
\forall\:y\!\cdot\!\exists\:z\!\cdot\!y\!*\!z\!>\!X_0
\end{array}\]
The two \emph{de Bruijn} terms are related structurally, but not semantically because of the
unmonitored shift of the context modifying free variables representation.

To address this problem and some others, numerous induction principles were derived in
\Bicoqd: (weak) structural induction, semantical induction, strong induction based on a
measure for a given type or for mutually recursive types. And this was not yet sufficient
for proof induction because the predecessors (sub-proofs) of a step in a proof have different
(dependent) types. This was not considered as a proper approach, because of the number of
principles to be expressed and proved as well as the absence of genericity of the proof
method.

For \Bicoqt a general approach has been designed. It combines a single induction principle
based on a measure in $\NAT$ (something rather intuitive) with a strategy for conducting the
proof defined through a inductive relation (so-called \emph{accessibility relation}). The
induction principle is generic, as $D$ is any family of types (indexed by $T$), $M$ any
measure and $P$ any predicate:
\[\small\begin{array}{l}
\forall\:(T\!:\!\Cmd{Type})(D\!:\!T\!\to\!\Cmd{Type})
         (M\!:\!\forall\:(t\!:\!T),\:D\:t\!\to\!\NAT)
         (P\!:\!\forall\:(t\!:\!T),\:D\:t\!\to\!\Cmd{Prop}),\\
(\forall\:(t\!:\!T)(d\!:\!D\:t),(\forall\:(t'\!:\!T)(d'\!:\!D\:t'),
 \:M\:t'\!<\!M\:t\!\to\:P\:t')\!\to\!P\:t)\!\to\!
\forall\:(t\!:\!T),\:P\:t
\end{array}\]
It does not describe what are the `smaller' terms to consider -- this results of the
selected accessibility relation. Choosing this relation is choosing the strategy, the cases in
a proof by cases, the predecessors for the entity you are considering. Intuitively, this
defines paths to reach terms in $D$, and provided the measure is compatible with the relation
(\emph{i.e.} predecessors are smaller) it allows to derive proofs along these paths. The
accessibility relation can be surjective or not in $D$; in the later case it defines a strict
subset of accessible terms and can be used to prove that any term of this subset satisfies a
property. For example a semantically relevant strategy can be defined as follows:
\[\small\begin{array}{l}
\Cmd{Inductive}\:\Sigma_{\text{Sem}}\!:\!\Btrm\!\to\!\Cmd{Type}\!:=\\
\:|\:\Sigma_\chi\!:\!\forall\:(i\!:\!\Bidx),\:\Sigma_{\text{Sem}}\:\evar{i}\\
\:|\:\Sigma_\forall\!:\!\forall\:(P\!:\!\Bprd)(i\!:\!\Bidx),\:
                  \Sigma_{\text{Sem}}\:P\!\to\!
                  \Sigma_{\text{Sem}}\:\dot{\forall}\:i\!\cdot\!P\\
\:|\:\Sigma_{\{\}}\!:\!\forall\:(P\!:\!\Bprd)(E\!:\!\Bexp)(i\!:\!\Bidx),\:
                  \Sigma_{\text{Sem}}\:P\!\to\!
                  \Sigma_{\text{Sem}}\:E\!\to\!
                  \Sigma_{\text{Sem}}\:\dot{\{}i\!:\!E\:\dot{|}\:P\dot{\}}\\
\:|\:\ldots\text{(straightforward induction)}\\
\end{array}\]
This relation is surjective, \emph{i.e.} $\forall\:(T\!:\!\Btrm),\:\Sigma_{\text{Sem}}(T)$. To
prove a property $Q$ for \emph{any} term $T$, it is possible to apply the generic induction
principle (with $M$ the standard depth function on \tool{B} terms) and then to use this
relation to make a proof by cases using inversion of the \tool{Coq} term
$\Sigma_{\text{Sem}}(T)$. The generated subgoals are then semantically relevant, \emph{e.g.}
$Q\:E'\!\to\!Q\:P'\!\to\!Q\:\dot{\{}i'\!:\!E'\:\dot{|}\:P'\dot{\}}$.

\subsection{About Lists, Maps and Abstract Data Types}\label{bicoq_lists}

Various syntactical entities are represented in our embedding, including sequents and parallel
substitutions (used as a technical tool to prove complex results presented thereafter). In
\Bicoqd these constructs are implemented through lists, but we have explored other
alternatives in \Bicoqt\:\!\!\!.

Proof environments in sequents are finite sets of predicates. In \Bicoqt they are represented
by a specification: signature of functions for membership, freeness, etc. with the appropriate
properties as axioms. The specification has the advantage to describe only what we \emph{need}
to know, and permits to use efficient concrete functions of the target language when there is
an implementation objective\footnote{E.g. \Bicoqt specifies the terms equality to use
\tool{OCaml}'s $=$ in the implementation.}. Yet we do \emph{not} recommend this approach for a
deep embedding, as the workload is not significantly reduced, whereas there is a risk to
introduce inconsistent axioms.

Another possibility adopted in \Bicoqt is the use of maps to represent parallel substitutions.
They can be described as lists of pairs in $\Bidx\!\times\!\Bexp$ provided that there are
never two pairs $(i,E)$ and $(i,E')$ s.t. $E\!\not=\!E'$, but it is more efficient to consider
them as functions in $\Bidx\!\to\!\Bexp$. In our experience, this approach simplifies the
development and the proofs -- consider the use of parallel substitutions to represent lifting:
it is not possible to build a generic lift substitution using finite lists, because any
index $i$ that can appear dangling in a term $T$ has to be modified, whereas a unique
(infinite) map can represent lifting for any term. On the other hand, maps require additional
theorems that may be complex to deal with as $\Bidx\!\to\!\Bexp$ is not well-founded. Yet the
main results consider parallel substitutions applied to a term, for which well-foundedness
holds. A more straightforward approach, yet to be explored, would be to reintroduce
well-foundedness through scoped maps, that is parallel substitutions represented by elements
of $(\Cmd{List}\:\Bidx)\!\times\!(\Bidx\!\to\!\Bexp)$, the list enumerating the relevant
indexes.

Maps are therefore efficient tools for deep embeddings, but our recommandation would be to
carefully analyse \emph{all} consequences of using such a design. For example, they cannot be
analysed extensionally -- just another way to say that they are not well-founded. That means
in practice \emph{e.g.} that as we need to be able to decide whether or not a variable appears
free in (one of the predicates of) a proof environment $\Gamma$, we cannot encode $\Gamma$ as
a function in $\Bprd\!\to\!\Bbol$. Indeed, being unable to identify \emph{a priori} predicates
of $\Gamma$, testing freeness would require examining \emph{all} predicates in the (infinite)
type $\Bprd$.

\subsection{Relationships between \emph{B} and \emph{Coq} logics}\label{bicoq_swap}

Deep embeddings such as \Bicoqd and \Bicoqt ensure a clear separation of the host and the
guest logics, allowing \emph{e.g.} for a study of their relations as illustrated here with the
\tool{B} operators on the left side and the \tool{Coq} operators on the right side:
\[\small\begin{array}{rcl}
\Gamma\Binf P_1 \pand P_2 & \quad\Leftrightarrow\quad &
 (\Gamma\Binf P_1)\land(\Gamma\Binf P_2)\\
\Gamma\Binf \dot{\forall}\:i\!\cdot\!P & \quad\Leftrightarrow\quad &
\forall\:(E\!:\!\Bexp),\:\Gamma\Binf\Baffec{i}{E}{P}\\
\Gamma\Binf P_1 \pimp P_2 & \Rightarrow & \Gamma\Binf P_1 \Rightarrow \Gamma\Binf P_2\\
\Gamma\Binf P_1 \por P_2 & \Leftarrow & (\Gamma\Binf P_1)\lor(\Gamma\Binf P_2)\\
\Gamma\Binf \dot{\exists}\:i\!\cdot\!P & \quad\Leftarrow\quad &
\exists\:(E\!:\!\Bexp),\:\Gamma\Binf\Baffec{i}{E}{P}\\
\Gamma\Binf E_1\pequ E_2 & \quad\Leftarrow\quad & E_1\!=\!E_2\\
\end{array}\]
The interesting results are those that are not equivalences. For example disjunction ($\lor$
vs $\por$) is very significant w.r.t. the difference between the classical logic of \tool{B}
and the constructive logic of \tool{Coq}. The \emph{excluded middle} being provable in
\tool{B}, it is always possible to provide a proof of $\Binf P\por\pnot P$; should the
disjunction being directly translated in \tool{Coq} we would obtain
$(\Binf P)\lor(\Binf\pnot P)$ for any $P$, that is a proof that the \tool{B} logic
is complete, which of course is not the case.

Note that these results provide a formal justification for the translation in a shallow
embedding; one may wonder whether it would be possible to automatically derive (or extract) a
shallow embedding from a deep embedding, provided such results.

\subsection{New Results and Enriched \emph{de Bruijn} Indexes}\label{bicoq_namespaces}

\subsubsection{Using Standard Indexes.} The \tool{B} inference rules defined in \cite{abr:1}
include a congruence rule: if $\Gamma\!\vdash\!E\!=\!F$ and $\Gamma\!\vdash\![x\!:=\!E]P$,
then $\Gamma\!\vdash\![x\!:=\!F]P$. \Bicoqd generalises this congruence rule to equivalent
predicates (extending the syntax with propositional variables). These results, however, are
limited to the replacement of \emph{unbound} subterms; that is, they are for example not
applicable to systematically simplify $\Gamma\Binf\dot{\forall}\:i\!\cdot\!(\pnot\pnot P)$
into $\Gamma\Binf\dot{\forall}\:i\!\cdot\! P$ as $i$ may appear free in $P$.

The substitution operator (left code) indeed mechanically avoid capture of variables by
enforcing lifting when crossing a binder. So \Bicoqt also addresses a more generic class of
congruence rules by defining \emph{grafting} (right code), which compared to the standard
substitution allows for the capture of variables in the parameter $E$ by \emph{never} lifting
it:
\[\small\begin{array}{lll}
 \begin{array}{l}
  \Baffec{i}{E}{}_d\!:\!\Btrm\!\to\!\Btrm\!:=\\
  \quad|\:\pfor T'\Rightarrow
          \pfor(\Baffec{\uparrow_d i}{\uparrow_d E}{}_{d\!+\!1} T')\\
  \quad|\:i'\Rightarrow \Cmd{if}\:i'\!=\!i\:\Cmd{then}\:E\:\Cmd{else}\:i'\\
  \quad|\:\ldots
 \end{array}
& \quad \quad \quad &
 \begin{array}{l}
  \Bgraftprd{i}{E}{}_d\!:\!\Btrm\!\to\!\Btrm\!:=\\
  \quad|\:\pfor T'\Rightarrow
        \pfor(\Bgraftprd{\uparrow_d i}{E}{}_{d\!+\!1} T')\\
  \quad|\:i'\Rightarrow \Cmd{if}\:i'\!=\!i\:\Cmd{then}\:E\:\Cmd{else}\:i'\\
  \quad|\:\ldots
 \end{array}
\end{array}\]
Grafting being defined, we have proven (using parallel substitutions) in \Bicoqt the following
congruence results for the replacement of sub-terms:
\[\small\begin{array}{lll}
\begin{array}{c}
\Binf E_1\pequ E_2\\\hline
\vspace{-9pt}\\
\Gamma\Binf\Bgraftprd{i}{E_1}{P}\piff\Bgraftprd{i}{E_2}{P}
\end{array}
& \quad \quad \quad &
\begin{array}{c}
\Binf E_1\pequ E_2\\\hline
\vspace{-9pt}\\
\Gamma\Binf\Bgraftprd{i}{E_1}{E}\pequ\Bgraftprd{i}{E_2}{E}
\end{array}
\end{array}\]
These results extend the classical congruence rules to bound subterms -- \emph{e.g.} they
justify why it is always valid to simplify a subterm $\lnot\lnot P$ into $P$, anywhere in a
term. But they are not generic enough, as the equality $E_1\!=\!E_2$ has to be proven in the
empty context. So they cannot for example be used to unfold a \emph{conditional} definition
such as $y\!\not=0\vdash x/y\!=\!\Cmd{max}\{z\!\in\!\NAT\:|\:y\!\times\!z\!\leq\!x\}$. This
limitation is not logical but technical. Preventing lifting when crossing a binder is
necessary to permit captures of variables, but causes a loss of context: free variables
representation is modified without control.

\subsubsection{Introducing Namespaces.} Several approaches were considered to avoid this
limitation of the congruence results: using names, marking \emph{De Bruijn} indexes during
grafting, defining grafting as the composition of primitive operations... to finally develop
for \Bicoqt a simpler solution, \emph{enriched de Bruijn indexes}.

In its most general form, this notation represents free and bound variables by pairs $(n,x)$,
the first parameter $n$ being the \emph{namespace} and the second one the index. Binders of
the language are themselves parameterised by a namespace in which they capture variables.
Namespaces can be seen as sorts, used to mark binders and indexes\footnote{Sorts for
\emph{de Bruijn} indexes are considered in \nocite{nad:1,nad:2}\cite{dar:1} but for different
reasons, each of the two binders of the defined language using its own space of \emph{de
Bruijn} indexes.}. This has limited consequences on the complexity of the code of the various
operations on terms, \emph{e.g.} lifting is as well parameterised by a namespace and only
modifies indexes in this namespace. This representation defines a form of names: if there is
no binder in a namespace $n$, a pair $(n,x)$ \emph{always} represents a free variable and can
be considered as a name, being \emph{never} subject to computations but dealt with using only
decidable equality.

\Bicoqt applies these principles in a simplified manner: the namespace set $\mathcal{N}$
contains (at least) two values, all the binders acting implicitly in the dedicated namespace
$n_0$, the other namespaces being used for eternally free variables. Consistently, lifting
only modifies pairs of the form $(n_0,x)$ in a term, etc. It is then possible to prove
improved congruence results:
\[\small\begin{array}{lll}
\begin{array}{c}
\Gamma\Binf E_1\pequ E_2 \quad
\Gamma\!\perp\! E_1\pequ E_2\\\hline
\vspace{-9pt}\\
\Gamma\Binf\Bgraftprd{i}{E_1}{P}\piff\Bgraftprd{i}{E_2}{P}
\end{array}
& \quad \quad \quad &
\begin{array}{c}
\Gamma\Binf E_1\pequ E_2 \quad
\Gamma\!\perp\! E_1\pequ E_2\\\hline
\vspace{-9pt}\\
\Gamma\Binf\Bgraftprd{i}{E_1}{E}\pequ\Bgraftprd{i}{E_2}{E}
\end{array}
\end{array}\]
The side condition $\perp$ requires $\Gamma$ and $E_1\!=\!E_2$ to have no common free
variable \emph{in the namespace} $n_0$ -- the technical difficulty is still there, but is now
limited to a dedicated namespace. Provided we avoid using the namespace $n_0$ for free
variables (through an extended form of $\alpha$-conversion, changing the name of the free
variables), we got the full expressiveness of our result, \emph{e.g.} allowing for the
replacement of conditional definitions. In their most general form, these results allow for
$\beta$-reduction, unfolding of (conditional) definitions, as well as the replacement
(rewriting) of equivalent subterms under a binder.

\subsubsection{Applicability of the New Results.} As noted in Sub. \ref{aboutb_motivations},
it is important to justify that such new results are truly applicable to \tool{B} and are not
artefacts provable only using features of the host logic. We provide the intuitive
justification by the \emph{Curry-Howard} isomorphim. The interpretation of the congruence
results is that provided a \tool{B} proof of $\Gamma\vdash E_1\!=\!E_2$, if
$\Gamma\!\perp\! E_1\pequ E_2$ then there \emph{always} exists a \tool{B} proof of
$\Gamma\vdash\Bgraftprd{i}{E_1}{P}\Leftrightarrow\Bgraftprd{i}{E_2}{P}$. In fact, the
\tool{Coq} proof \emph{is} a program building such a \tool{B} proof, the $\Sigma_{\text{Sem}}$
accessibility relation used in the \tool{Coq} proof (cf. Sub. \ref{bicoq_induction}) being the
recursion strategy of this program.

\section{Conclusion}\label{conclusion}

Through the presentation of two deep embeddings of the \tool{B} logic in \tool{Coq}, namely
\Bicoqd and \Bicoqt\:\!\!\!, we have discussed techniques to deal with deep embeddings, or
more generally with complex developments in higher-order logic (\tool{HOL}) frameworks --
\emph{e.g.} combining a generic induction scheme with \emph{ad hoc} accessibility relations or
implementing sets with maps rather than lists. One of these techniques applicable to language
mechanisations is to enrich \emph{de Bruijn} representation.

The first proposed adaptation enforces an explicit and precise management of the
$\lambda$-height parameter -- to the extent that it is added to operations that do not
strictly require it. This is in fact a form of encoding ensuring a consistent management of
the context: not only are proofs easier to conduct, but in some cases it also allows for finer
definitions and proofs of properties that would not be valid in a cruder version. Context
management is intuitive and don't require to use the full arithmetics: the only required
operators on indexes are successor, predecessor and comparison.

The second adaptation introduces \emph{namespaces} to parameterise binders and indexes. It is
a way to partition variables and to easily restrict scopes. Again, the required adaptations of
the operations are simple and intuitive, but the benefits are in our case important: beyond
obtaining the full power of complex congruence results, it is a frequent cause for proof
simplifications. Namespaces also define an approach to consider substitution and grafting as a
single operation: substitution is emulated by grafting provided free variables are in never
bound namespaces.

We may also note that our design choice is to directly encode application as an external
operation -- \emph{i.e.} a shallow representation of application in our deep embedding.
Together, these adaptations of the \emph{de Bruijn} representation seem to define a new form
of calculus for languages, of which detailed properties are still to be carefully studied and
compared to other calculi (\emph{e.g.} \cite{aba:1,cur:1,nad:1,nad:2}). Clearly, a full
version of this calculus easily represents the concept of sorts, provided with an efficient
management of contexts.

Taking the user view, these embeddings also demonstrate that it is possible to embed a non
trivial logic while ensuring accuracy and readability. Their usefulness to check the validity
of known results is illustrated by the identification of various oversights -- in our view a
sufficient justification for this activity, at least from a security perspective (cf.
\cite{jae:2}). The development of proven tools and the derivation of non trivial theorems that
were, in our knowledge, not proven in \tool{B} (without even speaking of formally checked) are
additional benefits.

\subsubsection{Acknowledgements} We thank Pr. C. Dubois for its advices.



\bibliographystyle{splncs}
\bibliography{BiCoq3}


\end{document}